# TOWARDS QUASI-BIOLOGICAL NANODOSIMETRY

M. Hajek

Institute of Atomic and Subatomic Physics, Vienna University of Technology, Austria

***Abstract***
*The increasing utilization of charged particle beams for therapeutic purposes requires designing novel detector systems which shall be capable of assessing radiation quality for a diversity of ion species. It is shown that the pattern of energy deposition in thermoluminescent phosphors and biological tissue contains conceptual parallels. The correlation of physical and radiobiological parameters observed experimentally for specific endpoints (single- and double-strand breaks of DNA) opens the possibility of realizing successfully quasi-biological solid-state nanodosimetry on the basis of thermoluminescence.*

## 1. Introduction

Significant progress in radiobiology has refined our understanding of radiation-induced biological response at the cellular level and challenged the conventional macroscopic description of radiation action in favour of a microdosimetric approach. It is inherent to the macroscopic concepts of absorbed dose and linear energy transfer (LET) that the energy deposition of a charged particle is treated as a continuous process along the particle track [1], neglecting the chaotic tangle of secondary electron paths of which it is composed. While the energy deposited by $\gamma$ rays at large doses can be imagined to be deposited 'homogeneously' relative to the size of biological targets, the energy deposited by heavy ions is much more heterogeneous. The universal use of dose as a normalizing parameter in radiobiology is based entirely on the availability of measuring instruments; it is a poor basis for predicting or understanding the relationship between an irradiation and the resulting endpoint [2]. This has stimulated the development of microdosimetry which takes into account the stochastic nature of interaction processes. But in the many years since its introduction, and in spite of the enormous efforts on its behalf, microdosimetry alone has not led to any fundamental understanding of radiobiology. Quoting Kellerer, "Concepts of microdosimetry are of course essential in any analysis of the action of ionizing radiation on the cell. Their employment has led to important insights but not, as yet, to a quantitative treatment of primary cellular changes" [3]. Up to now, the only instruments that are capable of measuring microdosimetric quantities, *e.g.* distributions of specific or lineal energy, are small gaseous proportional counters, scaled by density to cellular volumes of micrometre or nanometre diameter [4]. It would, however, be a great step forward to find a detector for which the pattern of energy-deposition events resembles the situation in a cell. Waligórski and Katz were the first to recognize that the response of solid-state thermoluminescence dosimeters (TLDs) depends on ionization density in a qualitatively similar way to the relative biological effectiveness (RBE) for many endpoints; this led them to conclude that TLDs appear to be "good candidates for mimicking the response of biological systems to heavy-ion irradiations" [5].

## 2. Microdosimetric multi-hit models for thermoluminescence

The occurrence of thermoluminescence (TL) is coupled to the presence of impurities or defects in a given substance. However, association of specific imperfections with a certain peak in the TL glow curve is not straightforward. It may very well happen that a certain impurity or defect is abundant in the sample, but does not contribute to the emitted TL. On the other hand, other defect structures, sometimes undetectable by other means due to their low concentrations, are found to be responsible for the TL signal [6]. From the physical point of view, understanding of the role of defect centres and the processes by which energy is first stored in the material and is then released in the form of light during heating of the sample is fundamental to develop new dosimeter materials with properties tailored to specific needs.

The most popular TL dosimeter in use today, LiF doped with Mg and Ti (most often containing additional OH impurities), has been available commercially since the late 1960s. The LiF crystal consists of two interpenetrating fcc lattices, one for $Li^+$ and one for $F^-$ ions. The ions are closely packed with a lattice constant of 0.4 nm. To describe the dose response and TL efficiency with respect to $^{60}$Co γ rays, Olko [7] has adopted microdosimetric multi-hit models which have originally been proposed to explain the inactivation of microorganisms. A solid-state TL detector contains a large number of independent structures (further on called 'targets'). It is assumed that only one type of target is present in the detector and each target can respond upon an energy deposit (termed a 'hit'). The target can tolerate a number of $m - 1$ hits without being affected; however, if $m$ or more hits occur, this will generate a response: TL emission in our case, cell inactivation or other endpoints in biological systems. Target size can be varied as a free parameter. The relative TL efficiency of the dominant peak 5 is found to decrease with ionization density (Fig. 1a) [8]. Its γ-ray response shows a linear-supralinear-sublinear slope. The related defect centre is assumed to be the combination of a 'one-hit' and 'two-hit' trap [5]. Being capable of capturing one charge carrier, a 'one-hit' trap requires a single energy deposit ($m = 1$) and produces a linear response which is found experimentally for modest doses <10 Gy. The characteristic target size of a 'one-hit' trap was estimated to be ~10 nm [5]. This model is confirmed by optical absorption measurements from which it is known that the structure responsible for peak 5 is composed of a $Mg^{2+}$-Li vacancy trimer (the electron trap) coupled to $Ti(OH)_n$ (the luminescence centre).

The relative efficiency of the high-temperature TL (HTTL, 248−310°C), on the other hand, increases with ionization density to pass through a maximum at an LET of ~100 keV/µm [8]. The slope is particularly impressive if the HTTL efficiency is scaled to the same level of absorbed dose (Fig. 1a); the resulting parameter is called the high-temperature ratio (HTR). This behaviour is a typical example of dominating 'two-hit' response ($m = 2$), the corresponding centre being capable of capturing two charge carriers of the same sign. Following high-LET irradiation, the 'two-hit' traps would be preferentially populated compared to traps associated with 'one-hit' centres, as the probability of multiple ionizations increases dramatically in vicinity of the particle track. The early onset of supralinear γ-ray dose response at ~200 mGy [9] further confirms the dominance of 'two-hit' centres being responsible for the HTTL; a pure 'two-hit' trap would produce a quadratic TL response. The characteristic target size of a 'two-hit' trap was estimated to be ~40 nm [5]. The molecular nature of the trapping and luminescence centres giving rise to the HTTL has not been identified yet. However, there is considerable evidence that Ti-related structures are involved [10].

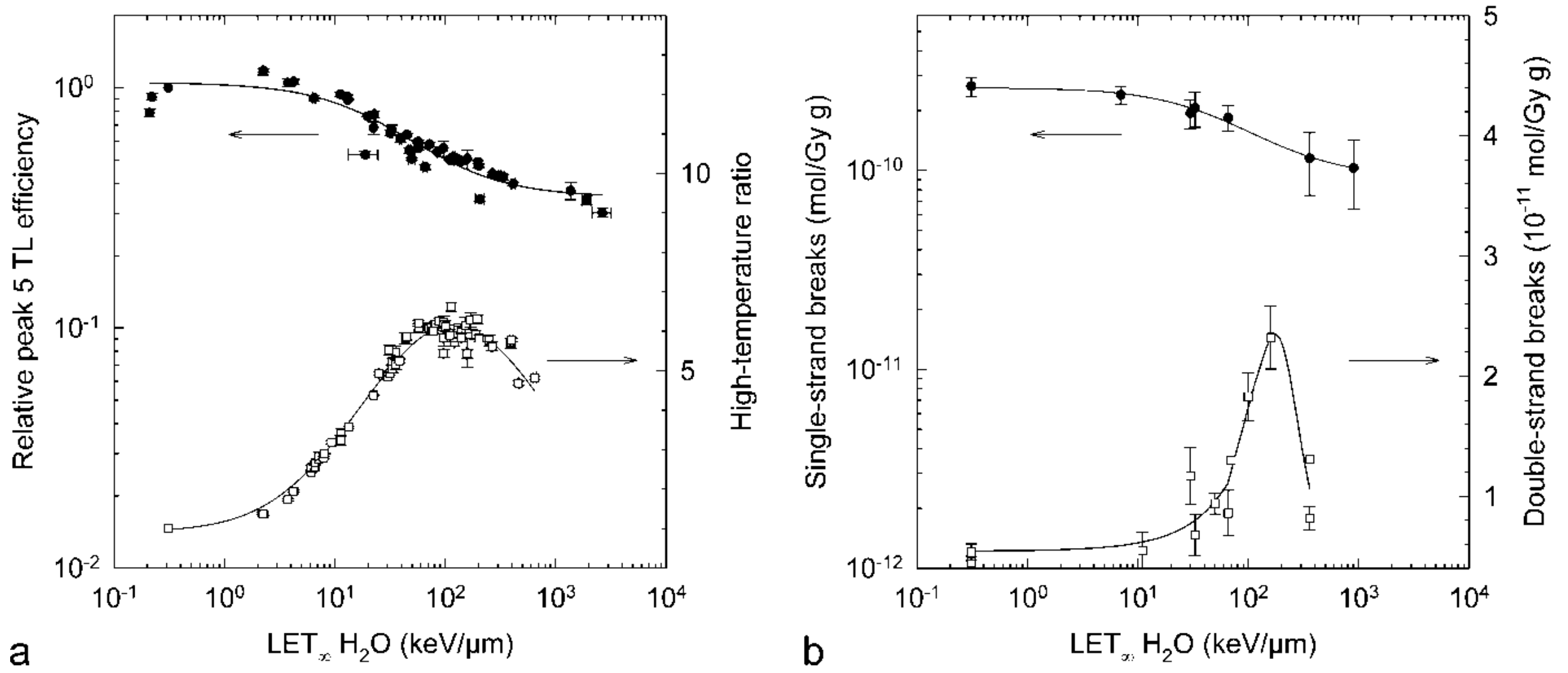

*Fig. 1: LET dependence of physical and biological response to 'one-hit' and 'two-hit' energy deposits[1)]. a, LiF:Mg,Ti peak 5 relative TL efficiency (●) and high-temperature ratio (□). b, Induction of single-strand breaks (●) and double-strand breaks (□) in Chinese hamster V79 cell DNA; numerical data from [11].*

## 3. Correlation of physical and radiobiological endpoints

In biological systems, the degree of irreversibility and/or functional lethality may be correlated with the distance between DNA single-strand breaks (SSBs). It has been estimated that double-strand breaks (DSBs) arise from SSBs within approximately ten base pairs, *i.e.* 3.4 nm [12]. SSBs are *per se* the consequence of a 'one-hit' response, while for DSBs the spatial correlation of two 'one-hit' events resulting in a 'two-hit' response is required. The yield of SSB and DSB induction, *i.e.* their efficiency, was measured by Kampf [11] for particles of different LET and effective charge. The slope of SSB induction (Fig. 1b) corresponds with the LiF:Mg,Ti peak 5 relative TL efficiency, while the dependence of the DSB yield on LET (Fig. 1b) may be correlated with the HTR. Fürweger *et al.* [13] exposed cultivated human skin fibroblasts and LiF:Mg,Ti TL detectors to high-energy $^{4}He^{2+}$, $^{12}C^{6+}$, $^{20}Ne^{10+}$, $^{28}Si^{14+}$ and $^{56}Fe^{26+}$ ions, with special emphasis being laid on the low-dose region of some ten mGy where bystander effects could be expected to contribute significantly to the overall radiation risk. The investigated biological effects included specific biochemical events that are known to play a major role in the early cellular response to DSBs, such as the formation of pATM (serine 1981), γH2AX (serine 139) and pDNA-PKcs (threonine 2609) foci. Analysis at three different points of time (20 min, 1 h, and 2 h) after irradiation could elucidate the time response of cellular signalling and damage repair. Again, the ionization density dependence of the initial radiation-induced DSB induction in both directly hit and bystander cells was found to be correlated to the HTTL.

1) LET is an insufficient parameter to characterize ionization density which depends on both LET and effective charge of the ionizing particle.

## 4. Conclusions

The correlation of physical and biological parameters observed experimentally for single- and double strand breaks of DNA points to conceptual analogies in the energy deposition mechanisms and gives reason to optimism that quasi-biological solid-state nanodosimetry on the basis of thermoluminescence can be realized in the short term. There are still some minor discrepancies, such as the actual sizes of the nanoscale targets of physical and biological structures, giving room for further improvement of the applied models.

## 5. References

[1] Kiefer, J.: Biological radiation effects. Berlin: Springer-Verlag, 1990

[2] R. Katz, Dose, Radiat. Res., 137(3), p. 410, 1994

[3] A. M. Kellerer, A generalized formulation of microdosimetric quantities, Radiat. Prot. Dosim., 31(1−4), p. 9, 1990

[4] H. H. Rossi, W. Rosenzweig, A device for the measurement of dose as a function of specific ionization, Radiology, 64(3), p. 404, 1955

[5] M. P. R. Waligórski, R. Katz, Supralinearity of peak 5 and peak 6 in TLD-700, Nucl. Instrum. Methods, 175(1), p. 48, 1980

[6] McKeever, S. W. S.: Thermoluminescence of solids. Cambridge: Cambridge University Press, 1985

[7] P. Olko, Microdosimetry, track structure and the response of thermoluminescence detectors, Radiat. Meas., 41(Suppl. 1), p. S57, 2007

[8] T. Berger, M. Hajek, TL-efficiency—Overview and experimental results over the years, Radiat. Meas., 43(2−6), p. 146, 2008

[9] T. Berger, M. Hajek, On the linearity of the high-temperature emission from $^{7}$LiF:Mg,Ti (TLD-700), Radiat. Meas., 43(9−10), p. 1467, 2008

[10] Horowitz, Y. (ed): Microdosimetric response of physical and biological systems to low- and high-LET radiations. Amsterdam: Elsevier, 2006

[11] G. Kampf, Induction of DNA double-strand breaks by ionizing radiation of different quality and their relevance for cell inactivation, Radiobiol. Radiother., 29(6), p. 631, 1988

[12] Rossi, H. H.; Zaider, M.: Microdosimetry and its applications. Berlin: Springer-Verlag, 1996

[13] C. Fürweger, M. Hajek, N. Vana, R. Kodym, R. Okayasu, Cellular signal transduction events as a function of linear energy transfer (LET), Radiat. Prot. Dosim., 126(1−4), p. 418, 2007